# Femtosecond signatures of optically induced magnons before ultrafast demagnetization

Reza Rouzegar[1], Oliver Franke[1], Gal Lemut[1], Oliver Gueckstock[1], Junwei Tong[1], Dieter Engel[2], Xianmin Zhang[3], Georg Woltersdorf[4], Piet Brouwer[1], Tobias Kampfrath[1], Quentin Remy[1]

[1]Department of Physics, Freie Universität Berlin, 14195 Berlin, Germany
[2]Max-Born-Institut für nichtlineare Optik und Kurzzeitspektroskopie, 12489 Berlin, Germany
[3]Key Laboratory for Anisotropy and Texture of Materials (Ministry of Education), School of Material Science and Engineering, Northeastern University, Shenyang 110819, China
[4]Institut für Physik, Martin-Luther-Universität Halle, 06120 Halle, Germany

## Abstract

Optically induced demagnetization of 3d metallic ferromagnets proceeds as fast as ~100 fs and is a crucial prerequisite for spintronic applications, such as ultrafast magnetization switching and spin transport. On the 100 fs time scale, the magnetization dynamics is widely understood in the context of temperature models considering energy transfers between conduction electrons, magnons and crystal lattice[1,2]. However, on even faster time scales, the flow of both angular momentum and energy between these subsystems has so far not been studied. Here, we measure ultrafast demagnetization by ultrabroadband THz-emission spectroscopy. We find that the rate of change of the magnetization does not rise instantaneously, but on a time scale as short as 10 fs. This rise is a signature that a transfer of angular momentum from the magnons to conduction electrons proceeds in less than 10 fs, before substantial demagnetization has happened. We further conclude that most of the spin dissipated by the lattice is transferred via magnon-lattice rather than electron-lattice interaction. These results show that the limiting speed of magnetization dynamics is not demagnetization, as generally believed, and harnessing the earliest magnon dynamics could be a new route towards an even faster spintronics.

# Figures

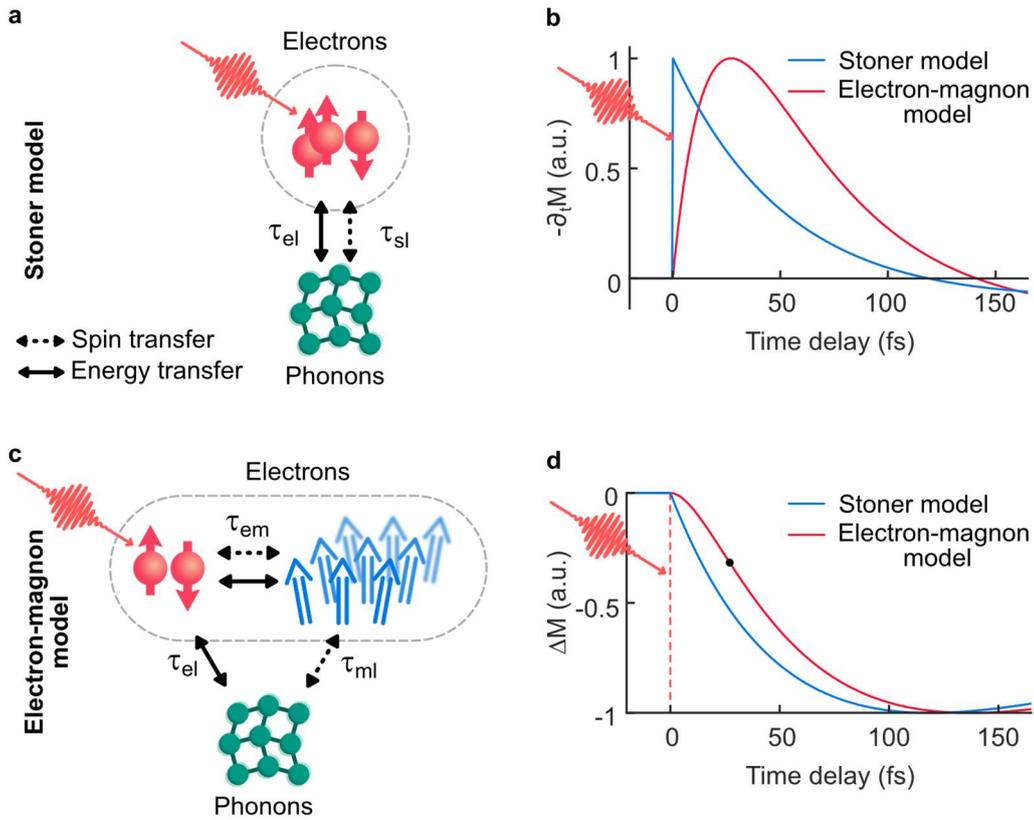

**Fig.1 | Models of UDM. a**, A femtosecond optical pulse heats the electrons of a metallic ferromagnet, resulting in magnetization quenching of the electron spins. In the Stoner model of ferromagnetism, the electron heating triggers transfer of energy (indicated by solid black arrow) and angular momentum (dashed arrow) between electrons and crystal lattice (phonons) via electron-lattice coupling and electron spin-flip scattering. The corresponding relaxation times are $\tau_{el}$ and $\tau_{sl}$. Because angular momentum transfer to the lattice starts instantaneously with electron heating at time $t = 0$, **b**, the $-\partial_t M$ rate of magnetization change (blue line) rises instantaneously. **c**, A more realistic model considers 2 types of spin excitations in the electronic subsystem: single-electron spin excitations (red arrows) and magnons (blue arrows). In simple terms, the additional relaxation time $\tau_{em}$ accounts for the generation of magnons while keeping the total electronic energy and angular momentum constant, which results in a non-instantaneous rise of $-\partial_t M$ (red line in **b**). This spin dissipation to the lattice happens from the magnons with relaxation time $\tau_{ml}$. **d**, Transient magnetization change $\Delta M(t)$ as expected from **b**.

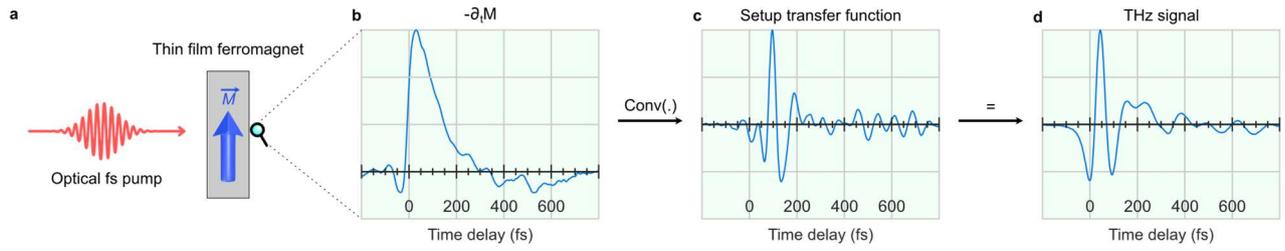

**Fig.2 | Experimental concept. a**, A ferromagnetic-metal layer FM made of, e.g., Py, with in-plane magnetization $M$, is excited by a 15 fs optical pump pulse. **b**, The resulting magnetization dynamics generates a THz electromagnetic pulse, whose electric field behind the sample scales with the time derivative $-\partial_t M$ of the magnetization (Supplementary Material). **c**, The THz electromagnetic pulse propagates through the experimental setup with the indicated transfer function and is detected by electro-optic sampling. **d**, Example of a measured THz signal, which is a convolution of the THz electric field (**b**) with the setup transfer function (**c**). In the experiment, we use FM|Pt stacks, where the Pt layer leads to a THz signal of larger amplitude but identical dynamics.

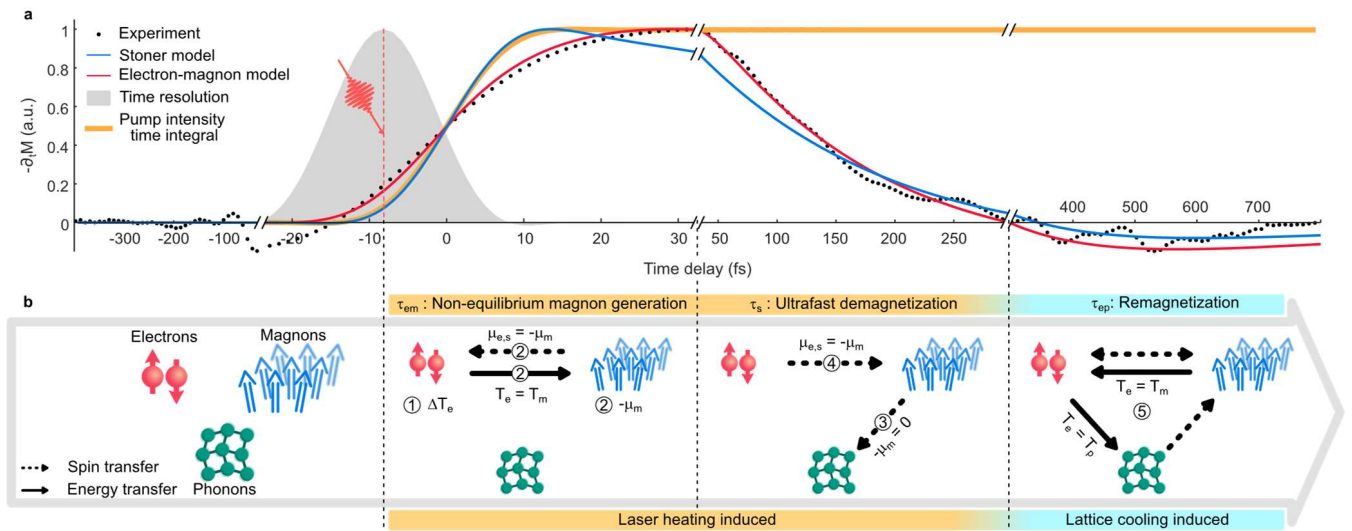

**Fig.3 | Dynamics of $-\partial_t M$ and interpretation. a,** Experimentally determined $-\partial_t M$ for Py (dotted line). The yellow line is the time integral of the effective pump-pulse intensity envelope (grey shaded area). The blue line is the fit using the Stoner model and the red line is the two-parameter fit by the electron-magnon model. The red dashed line indicates the position of the maximum of the laser pulse intensity in time according to the electron-magnon model. The yellow line was temporally shifted to provide the best fit of the initial dynamics. **b,** Proposed scenario of UDM in steps (1)-(5) within the electron-magnon model of Fig.1c. Solid and dashed arrows indicate transfer of energy and angular momentum, respectively. (1) Ultrafast electron heating by the pump pulse increases $T_e$ while keeping $T_m$ constant. (2) The difference $T_e - T_m$ triggers energy and spin transfer between the electrons and the magnons, resulting in an increase of $T_m$. Because of spin conservation in the combined electron-magnon subsystem, the electron spin accumulation $\mu_{e,s}$ increases, and the magnon chemical potential $\mu_m$ drops. Now, UDM is maximally efficient: (3) Magnons transfer spin to the lattice by magnon-lattice interaction at a rate $\propto \mu_m$. (4) Meanwhile, the resulting decrease of $-\mu_m$ causes a decrease of $\mu_{e,s}$ due to electron-magnon scattering. (5) After 300 fs, magnetization recovers by a combination of the reverse of steps (2) to (4) and cooling of the electrons by energy transfer to the lattice, analogous to "negative laser heating". The equations next to the spin/energy transfer arrows indicate the equilibrium condition governing each transfer.

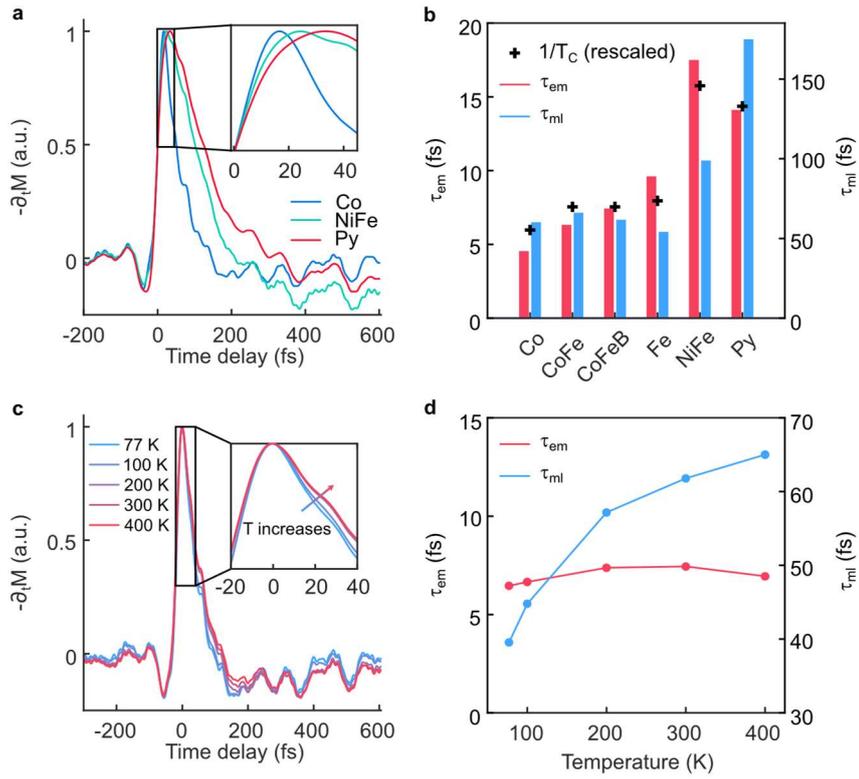

**Fig.4 | Impact of materials and temperature. a,** Measured rate $-\partial_t M$ of magnetization change vs time for different FM materials. The inset displays the same data for a shorter time range. **b,** Relaxation times $\tau_{em}$ and $\tau_{ml}$ as extracted from the data in **a** and the other materials (Supplementary Material) in conjunction with the electron-magnon model (Fig.1b). The inverse of the bulk Curie temperature[3–6], rescaled to the value of $\tau_{em}$ for Py, is also shown. **c,** Measured $-\partial_t M$ of CoFeB for different ambient temperatures and for CoFeB. **d,** Extracted relaxation times for the data in **c**.

**Table**

| Materials | Co | CoFe | CoFeB | Fe | NiFe | Py |
|---|---|---|---|---|---|---|
| $\tau_{em}$ (fs) | 4.5 | 6.3 | 7.4 | 9.6 | 17.5 | 14.1 |
| $\tau_{ml}$ (fs) | 60.2 | 66.2 | 61.8 | 54.2 | 98.9 | 175 |

**Table 1**: Extracted scattering times $\tau_{em}$ and $\tau_{ml}$ for each material.

## Introduction

In modern magnetism, ultrafast demagnetization (UDM) is a fundamental phenomenon that arises when a ferromagnetic material, such as a rare-earth[7] metal, 3d transition metal[8–12], semiconductor[13] or Heusler alloy[14], is excited by a femtosecond (fs) laser pulse. In the case of 3d transition metals, this effect is the fastest, with the magnetization amplitude $M$ dropping on an ultrafast time scale of 100 fs.

UDM has provided fundamental insights into the coupling of ordered spins to other degrees of freedom, such as the crystal lattice[15,16] and the electron orbital motion[17], and into the involved time scales[1,2]. From an applied perspective, UDM is highly interesting as an ultrafast source of angular momentum for transfer to an adjacent material[16,18–29], as done in magnetization switching[30–32] and the generation of THz electromagnetic pulses[33–39], or to the crystal lattice[15,16]. Consequently, it is important to develop an extensive understanding of UDM and the forces that drive it.

So far, UDM of single ferromagnets (FM) has been considered to proceed on time scales not faster than 100 fs. For example, in the framework of the Stoner model[23,40], which considers all electron degrees of freedom as one bath (Fig.1a), the rate of change $-\partial_t M$ of the magnetization $M$ with time $t$ rises step-like and subsequently decays with the time constant $(\tau_{sl}^{-1} + \tau_{el}^{-1})^{-1}$ (Fig.1b). Here, $\tau_{sl}$ and $\tau_{el}$ is the relaxation times at which the electron spins and the phonon system, respectively, adapts to the instantaneous rise of the electron temperature.

We expect that this situation changes profoundly if one considers two kinds of electronic spin excitations (Fig.1c), namely Stoner excitations (single-electron spin flips) and magnons (long-wavelength transverse spin excitations)[41]. In this more realistic view, the electronic degrees of freedom are accordingly separated into two baths, which we refer to as electrons (e) and magnons (m). Because the two baths can exchange both energy and angular momentum and the laser initially only deposits energy in the electrons, we expect a non-zero rise time $\tau_{em}$ of $-\partial_t M$ at $t = 0$ (Fig.1b)[41]. As a consequence, the change in magnetization $\Delta M(t)$ has an inflexion point and a delayed onset, in contrast to the instantaneous onset in the Stoner model (Fig.1d).

The existence of the electron and magnon baths is required to explain fundamental parameters of ferromagnetic metals, such as Curie temperature[42] and exchange splitting, in a realistic manner[3]. So far, however, no dynamic signatures of electron-magnon coupling were observed in UDM. Possible reasons are that such dynamics are potentially significantly faster than the typical time resolution of 50 fs used to measure UDM[2] (see Supplementary Material) and that most experiments cannot reveal the inflexion point of $\Delta M(t)$ (Fig.1c) even for a time resolution down to 30 fs (see Supplementary Material).

In this work, we use THz-emission spectroscopy[23] with an ultrawide bandwidth of 1-65 THz to obtain a direct and faithful access to the pump-induced dynamics of $-\partial_t M$ with a time resolution of ~15 fs. We find a rise time in a broad range of 3d ferromagnets, ranging from 5 fs in Co to 18 fs in $Ni_{50}Fe_{50}$. This feature is more than one order of magnitude shorter than the subsequent UDM found in ferromagnets. We ascribe the ultrafast rise of $-\partial_t M$ to the generation of non-equilibrium magnons[41,43–46] due to the transfer of both energy and angular momentum between electrons and magnons, happening before UDM sets in (Fig.1c).

## Experimental concept

In our experiment, we excite FM films with an optical pump pulse and measure the resulting UDM by suitable probes (Fig.2a). As samples, we study a broad range of ferromagnetic 3d transition metals and their alloys, i.e., Co, $Co_{70}Fe_{30}$ (CoFe), $Co_{60}Fe_{20}B_{20}$ (CoFeB), Fe, $Ni_{50}Fe_{50}$ (NiFe) and permalloy $Ni_{80}Fe_{20}$ (Py).

To identify UDM features significantly faster than 50 fs, the following conditions have to be addressed: (i) We need to measure $-\partial_t M$ rather than $\Delta M(t)$ to pronounce very fast features (see Fig.1b vs 1d). (ii) A faithful probe of $-\partial_t M$ is needed, without the presence of other signal components. (iii) We need to achieve a time resolution significantly finer than 50 fs.

The requirements (i) and (ii) are almost ideally fulfilled by THz emission spectroscopy: The pump-induced $\partial_t M$ acts like a magnetic dipole[47–49] and emits an electromagnetic pulse with transient electric field $E_{\partial_t M}(t) \propto \partial_t M$ directly behind the sample (see Fig.2a,b). To enhance the relatively weak THz fields, we use FM|Pt stacks[23], where UDM is accompanied by a spin current $j_S(t)$ from FM to Pt. In Pt, the $j_S(t)$ is converted into an in-plane charge current that acts like an electric dipole[33–39] and, thus, also emits a THz pulse with field $E_{j_S}(t) \propto j_S(t)$ right behind the sample. We checked explicitly that $E_{\partial_t M}(t)$ from FM and $E_{j_S}(t)$ from FM|Pt have precisely the same dynamics (see Supplementary Material). To fulfill requirement (iii), we use pump pulses with a duration as short a ~15 fs from a Ti:sapphire laser oscillator and monitor the emitted the THz pulses by ultrabroadband electro-optic sampling (EOS)[50–52] with ~15 fs laser pulses in a 10 μm thin ZnTe crystal.

The measured THz signals result from the propagation of the THz electromagnetic pulse to the far-field, toward the ZnTe detection crystal, and subsequent EOS (see Fig.2). The THz field behind the sample (see Fig.2b) is, therefore, recovered from the EOS signal (see Fig.2d) by deconvolution of our setup transfer function (see Fig.2c)[48,53]. The latter is obtained via a proper ultrabroadband calibration measurement with a reference emitter (see Supplementary Material).

**Results**

**Rise time of $-\partial_t M$ and Stoner model**

Fig.3a shows the measured $-\partial_t M$ for Py (dotted black line), which exhibits a rise and subsequent decay on a time scale of 20 fs and 100 fs, respectively. The effective intensity profile $I_p(t)$ of the pump pulse is given by the gray shaded curve, whose full width at half maximum (FWHM) of 15 fs quantifies the time resolution of $-\partial_t M$ (Supplementary Material).

Let us assume for the moment that the FM response is instantaneous and long-lived on the time scale of the covered time window of ~1 ps (Fig.3a). In this case, $-(\partial_t M)(t)$ would be proportional to the Heaviside step function $\Theta(t)$. As the system response is linear (Supplementary Material) with respect to the perturbing pump intensity $I_p(t)$, the actual response $-(\partial_t M)(t)$ would scale with the convolution $(\Theta * I_p)(t) = \int_{-\infty}^{t} dt'\, I_p(t')$, i.e., the time integral of $I_p(t)$. Strikingly, the resulting curve rises substantially faster than the measured $-\partial_t M$ (Fig.3a).

Next, we use the Stoner model (Supplementary Material and Fig.1a) to fit the measured $-\partial_t M$. The best fit is obtained for time constants of $\tau_{sl} = 250$ fs and $\tau_{ep} = 267$ fs. While good agreement of experiment and model is obtained for $t > 100$ fs, the model is not able to reproduce the dynamics in the vicinity of the pump pulse between $-10$ fs and 100 fs (Fig.3a). This shortcoming is conceivable because, with a single electronic bath, the excess of spin to be dissipated in the lattice can only be generated from the direct interaction between electrons and the laser pulse. Therefore, in the Stoner model, the response $-\partial_t M$ to an infinitely short pump pulse is initially step-like (Fig.1d).

We conclude that the observed rate of change $-\partial_t M$ of the Py magnetization contains clear signatures of a non-instantaneous rise on a time scale of 10 fs that cannot be described by the Stoner model.

**The electron-magnon model**

To explain our experimental finding, we consider the more realistic and quite general electron-magnon model of a ferromagnet (Fig.1c)[41,43,44,54–58], where the electronic system is split in two subsystems, magnons and electrons[59]. We are interested in the flow of energy and angular momentum between all subsystems of electrons (e), magnons (m) and phonons (p). As a minimum set of conjugate thermodynamic variables, we use the electron temperature $T_e$, the electron spin voltage $\mu_{e,s}$, the magnon chemical potential $\mu_m$, the magnon temperature $T_m$ and the phonon temperature $T_p$ (Supplementary Material). Before arrival of the pump pulse, all temperatures are equal to 300 K, and all potentials are zero.

We consider electron-magnon (em), magnon-lattice[60] (ml) and electron-lattice[61,62] (el) scattering as the dominant mechanisms. These interactions are considered phenomenologically in the relaxation-time approximation. By taking advantage of conservation laws relevant to each interaction, only three characteristic times $\tau_{em}$, $\tau_{ml}$ and $\tau_{el}$ remain, which describe, respectively, the equilibration of energy and angular momentum between electrons and magnons, angular momentum between magnons and the lattice, and energy between electrons and the lattice (Fig.1c).

Fitting of the measured $-\partial_t M$ yields $\tau_{em} = 14$ fs and $\tau_{ml} = 175$ fs (see Figs.3a, 4a-b and Table 1), while the other parameters are fixed by experiments and *ab initio* calculations (Supplementary Material). A similar high fit quality is also obtained for other ferromagnetic materials and various ambient temperatures (Fig.4). The excellent agreement of the measured and modeled $-\partial_t M$ (Fig.3a) suggests that our very general rate-equation model captures the relevant physics of the ferromagnet. The model and its agreement with experiments allow us to extract the following main steps (1)-(5) of UDM (see Fig.3b).

(1) Optical heating of the electrons leads to a temperature imbalance with the magnons and, thus, (2) drives energy and spin transfer from electrons to magnons via electron-magnon scattering, resulting in an increase of $T_m$ and a simultaneous increase of $\mu_{e,s}$ and $-\mu_m$, because electron-magnon scattering preserves the combined spin of the electron and magnon systems. The reason that the temperature difference $T_e - T_m$ also drives changes in $\mu_{e,s}$ and $-\mu_m$ is that increasing the magnon energy via magnon creation automatically leads to a (negative) increase of the magnon angular momentum and a (positive) increase of the electron spin.

(3) Now, due to $\mu_m \neq 0$, spin is transferred from the magnons to the lattice, and the total angular momentum of magnons and electrons decreases. (4) Due to the strong electron-magnon coupling, the spin voltage follows the magnon chemical potential quasi-instantaneously (Supplementary Material). Finally, after around 300 fs, $\mu_{e,s}$ and $\mu_m$ reach zero, and the demagnetization stops. (5) Because electron-lattice interaction leads to electron cooling and a reduction of $T_e$ below $T_m$, the process reverses, and the sample re-magnetizes ($-\partial_t M < 0$).

Importantly, the step (2) leads to a delay of the actual UDM that takes place only in step (3). Note that the processes (1)-(4) do not happen sequentially but rather simultaneously (Supplementary Material). Importantly, a nonvanishing $\tau_{em}$ can lead to a nonvanishing rise time of $-\partial_t M$. The reason is that $-\partial_t M$ is proportional to $\mu_m$ and that electron-magnon scattering leads to a gradual build-up of $\mu_m$ through spin and energy transfer between electrons and magnons. For $\tau_{em} = 0$ fs,[54] the Stoner result is recovered, which considers an instantaneous exchange of energy and angular momentum between electron and magnon subsystems.

In the Supplementary Material, we discuss alternative dominant mechanisms in our electron-magnon model and find that they are not sufficient to describe our experimental results or even contradict them.

**Material dependence**

We conduct analogous experiments with more FM materials and find qualitatively identical dynamics for all tested 3d transition metals (Fig.3a and Supplementary Material). We observe a different rise time for different materials, with the maximum of $-\partial_t M$ reached in around 16 fs for Co up to around 33 fs for Py. Interestingly, the decaying dynamics follows a similar trend, as in part observed previously[23]. For a more quantitative analysis, we fit the dynamics for all materials using the electron-magnon model. The corresponding relaxation times $\tau_{em}$ and $\tau_{ml}$ are summarized in Fig.3b and Table 1. These results show that the speed of UDM ($\tau_{ml}$) and magnon generation dynamics ($\tau_{em}$, see inset of Fig.3a) vary for different materials. The extracted $\tau_{em}$ is as low as 4.5 fs in Co. We note that the time at which the $-\partial_t M$ maximum is reached is a combination of both $\tau_{em}$ and $\tau_{ml}$. Thus, Py has a smaller magnon generation time than NiFe, even though its $-\partial_t M$ peak amplitude is reached for a longer time delay.

Remarkably, Fig.3b suggests a correlation of the electron-magnon relaxation time $\tau_{em}$ with the inverse of the Curie temperature of the FM. This observation is plausible since both quantities are governed by magnetic exchange[42]. A more detailed understanding for such a correlation requires a microscopic treatment of electron-magnon scattering (Supplementary Material), as provided in Refs. [54] and [58]. Indeed, both treatments yield a $\tau_{em}$ proportional to the inverse of the Curie temperature. The correlation is less pronounced for the magnon-lattice equilibration time $\tau_{ml}$.

**Temperature dependence**

We finally measure $-\partial_t M$ of CoFeB as a function of temperature (Fig.3c). From 77 K up to 400 K, we notice a clear change in the decaying dynamics while the rising dynamics appears mostly unchanged. The corresponding relaxation times obtained from our electron-magnon model are shown in Fig.3d. Indeed, we find that the magnon-lattice relaxation time $\tau_{ml}$ increases with temperature. This behavior is a manifestation of the critical slowing down[63–66] of magnetization dynamics as the FM approaches the Curie temperature. Surprisingly, however, the electron-magnon relaxation time $\tau_{em}$ is found to be relatively temperature-independent. These features can also be reproduced by microscopic models[54,58] (see Supplementary Material). In the Supplementary Material, we discuss how the weak temperature dependence of $\tau_{em}$ can also arise from the energy dependence of the electronic density of states.

## Discussion

Our observation of a non-instantaneous onset of UDM, combined with our semi-phenomenological modelling, implies the generation of non-equilibrium magnons at the early stages of UDM. This finding allows us to draw a new picture of ultrafast magnetization dynamics as a whole. Interestingly, we find that, contrary to what was assumed for most models of UDM with electron-magnon scattering[27,41,44,54,55], spin dissipation from the conduction electrons to the lattice is not sufficiently efficient to explain UDM. This property is connected to a realistic magnon dispersion relation, which leads to a large magnon spin susceptibility, and the experimental requirement of a short electron-magnon scattering time (see Supplementary Material). The angular momentum dissipated in the lattice, therefore, stems from the magnons through magnon-lattice scattering. Because the magnon bath loses angular momentum via this process, it is only allowed via spin-orbit coupling (SOC)[67].

From the fits, we find that $\tau_{em}$ is significantly smaller than $\tau_{ml}$, in agreement with the fact that the exchange interaction leads to a faster dynamics than SOC[68]. The value of $\tau_{em}$ for iron and cobalt (Table I) agrees well with the lifetime of high-energy magnons[45], which are believed to contribute the most to UDM due to their high density of states[41,43]. Likewise, these extremely short times are consistent with the values of energy-dependent lifetimes of hot minority-spin electrons, which are found to be strongly impacted by the presence of magnons at ultrashort time scales[46].

Tengdin et al.[69] found indications of a sub-20 fs transfer of energy to magnons in nickel, by monitoring the maximum electron temperature as a function of fluence. This energy transfer is consistent with our electron-magnon model, and we observe it in all the studied metallic ferromagnets. However, a transfer of energy to the magnons does not necessarily imply the generation of magnons, but could also signal a redistribution of magnons toward higher energies only. In our low-fluence experiments, this process corresponds to the appearance of a nonzero magnon chemical potential. The generation of magnons requires a transfer of angular momentum, which at the sub-10 fs scale happens between electrons and magnons[41].

## Conclusion

We find that ultrafast spin dynamics already starts before UDM in less than 10 fs for most 3d transition metal, down to less than 5 fs for cobalt, via the generation of non-equilibrium magnons and simultaneous transfer of angular momentum to single-electron excitations. This newly discovered effect, at such short time scales, involves both energy and spin transfer, and can, thus, not be described by conventional temperature models[70,71].

Our findings should motivate further theoretical works regarding both electron-magnon scattering and its temperature dependence, as well as a non-equilibrium microscopic description of angular momentum exchange in ml scattering and its connection to critical slowing-down, which has never been tackled to the best of our knowledge. These concepts may also reveal undiscovered ultrafast processes in more exotic systems relying on a strong exchange interaction between different spin degrees of freedom, such as in heavy-fermion systems[72], heavy-metal-based antiferromagnetic alloys[73] and triplet superconductors[74].

## Acknowledgments

We thank Prof. Dr. Martin Wolf and the Max Planck Society for their strong support, as part of these measurements were conducted at the Fritz Haber Institute of the Max Planck Society. We also thank Liane Brandt for contributions at an early stage of this work. We gratefully acknowledge financial support from the German research foundation (DFG) through the collaborative research center CRC/TRR 227 (project ID 328545488, projects A05, B02, B03), the ERC-2023 Advanced Grant ORBITERA (grant no. 101142285) and the ERC Proof of Concept Grant T-SPINDEX (grant no. 101123255). Q. Remy acknowledges guiding discussions with Alexander Chekhov and Hans Christian Schneider and support by the Alexander von Humboldt Foundation

## Competing interests

Authors declare no conflict of interests.

## References

1.  Scheid, P., Remy, Q., Lebègue, S., Malinowski, G. & Mangin, S. Light induced ultrafast magnetization dynamics in metallic compounds. *J. Magn. Magn. Mater.* **560**, 169596 (2022).

2.  Chen, X. *et al.* Ultrafast demagnetization in ferromagnetic materials: Origins and progress. *Phys. Rep.* **1102**, 1–63 (2025).

3.  Kübler, J. *Theory of Itinerant Electron Magnetism*. (Oxford University Press, 2009).

4.  Sato, H. *et al.* Temperature-dependent properties of CoFeB/MgO thin films: Experiments versus simulations. *Phys. Rev. B* **98**, 214428 (2018).

5.  Yu, P., Jin, X. F., Kudrnovský, J., Wang, D. S. & Bruno, P. Curie temperatures of fcc and bcc nickel and permalloy: Supercell and Green's function methods. (2008) doi:10.1103/PhysRevB.77.054431.

6.  Milone, A. F., Ortalli, I. & Soardo, G. P. Curie temperature of Ni−Fe alloys in the region (24÷35)% Ni from Mössbauer experiments. *Nuovo Cim. D* **1**, 18–20 (1982).

7.  Wietstruk, M. *et al.* Hot-Electron-Driven Enhancement of Spin-Lattice Coupling in Gd and Tb 4f Ferromagnets Observed by Femtosecond X-Ray Magnetic Circular Dichroism. *Phys. Rev. Lett.* **106**, 127401 (2011).

8.  Beaurepaire, E., Merle, J.-C., Daunois, A. & Bigot, J.-Y. Ultrafast Spin Dynamics in Ferromagnetic Nickel. *Phys. Rev. Lett.* **76**, 4250–4253 (1996).

9.  Hohlfeld, J., Matthias, E., Knorren, R. & Bennemann, K. H. Nonequilibrium Magnetization Dynamics of Nickel. *Phys. Rev. Lett.* **78**, 4861–4864 (1997).

10. Scholl, A., Baumgarten, L., Jacquemin, R. & Eberhardt, W. Ultrafast Spin Dynamics of Ferromagnetic Thin Films Observed by fs Spin-Resolved Two-Photon Photoemission. *Phys. Rev. Lett.* **79**, 5146–5149 (1997).

11. Aeschlimann, M. *et al.* Ultrafast Spin-Dependent Electron Dynamics in fcc Co. *Phys. Rev. Lett.* **79**, 5158–5161 (1997).

12. Conrad, U., Güdde, J., Jähnke, V. & Matthias, E. Ultrafast electron and magnetization dynamics of thin Ni and Co films on Cu(001) observed by time-resolved SHG. *Appl. Phys. B Lasers Opt.* **68**, 511–517 (1999).

13. Wang, J. *et al.* Ultrafast Quenching of Ferromagnetism in InMnAs Induced by Intense Laser Irradiation. *Phys. Rev. Lett.* **95**, 167401 (2005).

14. Zhang, W. *et al.* Ultrafast demagnetization of Co2MnSi1-xAlx Heusler compounds using terahertz and infrared light. *Phys. Rev. B* **107**, 224408 (2023).

15. Dornes, C. *et al.* The ultrafast Einstein–de Haas effect. *Nature* **565**, 209–212 (2019).

16. Tauchert, S. R. *et al.* Polarized phonons carry angular momentum in ultrafast demagnetization. *Nat. |* **602**, 73 (2022).

17. Seifert, T. S. *et al.* Time-domain observation of ballistic orbital-angular-momentum currents with giant relaxation length in tungsten. *Nat. Nanotechnol.* **18**, 1132–1138 (2023).

18. Battiato, M., Carva, K. & Oppeneer, P. M. Superdiffusive Spin Transport as a Mechanism of Ultrafast Demagnetization. *Phys. Rev. Lett.* **105**, 027203 (2010).

19. Eschenlohr, A. *et al.* Ultrafast spin transport as key to femtosecond demagnetization. *Nat. Mater.*

**12**, 332–336 (2013).

20. Rudolf, D. *et al.* Ultrafast magnetization enhancement in metallic multilayers driven by superdiffusive spin current. *Nat. Commun.* **3**, 1037 (2012).

21. Malinowski, G. *et al.* Control of speed and efficiency of ultrafast demagnetization by direct transfer of spin angular momentum. *Nat. Phys.* **4**, 855–858 (2008).

22. Beens, M., de Mare, K. A., Duine, R. A. & Koopmans, B. Spin-polarized hot electron transport versus spin pumping mediated by local heating. *J. Phys. Condens. Matter* **35**, 035803 (2023).

23. Rouzegar, R. *et al.* Laser-induced terahertz spin transport in magnetic nanostructures arises from the same force as ultrafast demagnetization. *Phys. Rev. B* **106**, 144427 (2022).

24. Géneaux, R. *et al.* Spin Dynamics across Metallic Layers on the Few-Femtosecond Timescale. *Phys. Rev. Lett.* **133**, (2024).

25. Chen, J. *et al.* Competing Spin Transfer and Dissipation at Co/Cu(001) Interfaces on Femtosecond Timescales. *Phys. Rev. Lett.* **122**, 067202 (2019).

26. Shokeen, V. *et al.* Spin Flips versus Spin Transport in Nonthermal Electrons Excited by Ultrashort Optical Pulses in Transition Metals. *Phys. Rev. Lett.* **119**, 107203 (2017).

27. Beens, M., Duine, R. A. & Koopmans, B. Modeling ultrafast demagnetization and spin transport: The interplay of spin-polarized electrons and thermal magnons. *Phys. Rev. B* **105**, 144420 (2022).

28. Razdolski, I. *et al.* Nanoscale interface confinement of ultrafast spin transfer torque driving non-uniform spin dynamics. *Nat. Commun.* **8**, 15007 (2017).

29. Turgut, E. *et al.* Controlling the Competition between Optically Induced Ultrafast Spin-Flip Scattering and Spin Transport in Magnetic Multilayers. *Phys. Rev. Lett.* **110**, 197201 (2013).

30. Iihama, S. *et al.* Single-Shot Multi-Level All-Optical Magnetization Switching Mediated by Spin Transport. *Adv. Mater.* **30**, 1804004 (2018).

31. Remy, Q. *et al.* Accelerating ultrafast magnetization reversal by non-local spin transfer. *Nat. Commun.* **14**, 445 (2023).

32. Igarashi, J. *et al.* Optically induced ultrafast magnetization switching in ferromagnetic spin valves. *Nat. Mater.* **22**, 725–730 (2023).

33. Kampfrath, T. *et al.* Terahertz spin current pulses controlled by magnetic heterostructures. *Nat. Nanotechnol.* **8**, 256–260 (2013).

34. Seifert, T. *et al.* Efficient metallic spintronic emitters of ultrabroadband terahertz radiation. *Nat. Photonics* **10**, 483–488 (2016).

35. Yang, D. *et al.* Powerful and Tunable THz Emitters Based on the Fe/Pt Magnetic Heterostructure. *Adv. Opt. Mater.* **4**, 1944–1949 (2016).

36. Wu, Y. *et al.* High-Performance THz Emitters Based on Ferromagnetic/Nonmagnetic Heterostructures. *Adv. Mater.* **29**, (2017).

37. Torosyan, G., Keller, S., Scheuer, L., Beigang, R. & Papaioannou, E. T. Optimized Spintronic Terahertz Emitters Based on Epitaxial Grown Fe/Pt Layer Structures. *Sci. Rep.* **8**, 1311 (2018).

38. Schneider, R. *et al.* Magnetic-Field-Dependent THz Emission of Spintronic TbFe/Pt Layers. *ACS Photonics* **5**, 3936–3942 (2018).

39. Jungfleisch, M. B. *et al.* Control of Terahertz Emission by Ultrafast Spin-Charge Current

Conversion at Rashba Interfaces. *Phys. Rev. Lett.* **120**, 207207 (2018).

40. Nolting, W. & Ramakanth, A. *Quantum Theory of Magnetism*. (Springer Berlin Heidelberg, Berlin, Heidelberg, 2009). doi:10.1007/978-3-540-85416-6.

41. Dusabirane, F., Leckron, K., Rethfeld, B. & Schneider, H. C. Interplay of Electron-Magnon Scattering and Spin-Orbit Induced Electronic Spin-Flip Scattering in a two-band Stoner model.

42. Pajda, M., Kudrnovský, J., Turek, I., Drchal, V. & Bruno, P. Ab initio calculations of exchange interactions, spin-wave stiffness constants, and Curie temperatures of Fe, Co, and Ni. *Phys. Rev. B* **64**, 174402 (2001).

43. Weißenhofer, M. & Oppeneer, P. M. Ultrafast Demagnetization Through Femtosecond Generation of Non-Thermal Magnons. *Adv. Phys. Res.* (2024) doi:10.1002/apxr.202300103.

44. Tveten, E. G., Brataas, A. & Tserkovnyak, Y. Electron-magnon scattering in magnetic heterostructures far out of equilibrium. *Phys. Rev. B* **92**, 180412 (2015).

45. Zhang, Y., Chuang, T. H., Zakeri, K. & Kirschner, J. Relaxation time of terahertz magnons excited at ferromagnetic surfaces. *Phys. Rev. Lett.* **109**, (2012).

46. Schmidt, A. B. *et al.* Ultrafast magnon generation in an Fe film on Cu(100). *Phys. Rev. Lett.* **105**, (2010).

47. Huisman, T. J., Mikhaylovskiy, R. V, Tsukamoto, A., Rasing, T. & Kimel, A. V. Simultaneous measurements of terahertz emission and magneto-optical Kerr effect for resolving ultrafast laser-induced demagnetization dynamics. *Phys. Rev. B* **92**, 104419 (2015).

48. Zhang, W. *et al.* Ultrafast terahertz magnetometry. *Nat. Commun.* **11**, 4247 (2020).

49. Beaurepaire, E. *et al.* Coherent terahertz emission from ferromagnetic films excited by femtosecond laser pulses. *Appl. Phys. Lett.* **84**, 3465–3467 (2004).

50. Wu, Q. & Zhang, X.-C. Free-space electro-optics sampling of mid-infrared pulses. *Appl. Phys. Lett.* **71**, 1285–1286 (1997).

51. Leitenstorfer, A., Hunsche, S., Shah, J., Nuss, M. C. & Knox, W. H. Detectors and sources for ultrabroadband electro-optic sampling: Experiment and theory. *Appl. Phys. Lett.* **74**, 1516–1518 (1999).

52. Kampfrath, T., Nötzold, J. & Wolf, M. Sampling of broadband terahertz pulses with thick electro-optic crystals. *Appl. Phys. Lett.* **90**, 231113 (2007).

53. Seifert, T. S. *et al.* Femtosecond formation dynamics of the spin Seebeck effect revealed by terahertz spectroscopy. *Nat. Commun.* **9**, 2899 (2018).

54. Beens, M., Duine, R. A. & Koopmans, B. S-d model for local and nonlocal spin dynamics in laser-excited magnetic heterostructures. *Phys. Rev. B* **102**, 054442 (2020).

55. Cywiński, Ł. & Sham, L. J. Ultrafast demagnetization in the sp-d model: A theoretical study. *Phys. Rev. B* **76**, 045205 (2007).

56. Haag, M., Illg, C. & Fähnle, M. Role of electron-magnon scatterings in ultrafast demagnetization. *Phys. Rev. B* **90**, 014417 (2014).

57. Kang, K. & Choi, G. M. Thermal coupling parameters between electron, phonon, and magnon of Nickel. *J. Magn. Magn. Mater.* **514**, (2020).

58. Manchon, A., Li, Q., Xu, L. & Zhang, S. Theory of laser-induced demagnetization at high

temperatures. *Phys. Rev. B* **85**, 064408 (2012).

59. Izuyama, T. Collective Excitations of Electrons in Degenerate Bands. I. *Prog. Theor. Phys.* **23**, 969–983 (1960).

60. Schmidt, R., Wilken, F., Nunner, T. S. & Brouwer, P. W. Boltzmann approach to the longitudinal spin Seebeck effect. *Phys. Rev. B* **98**, 134421 (2018).

61. Grimvall, G. *The Electron-Phonon Interaction in Metals*. (North-Holland Publishing Co., Amsterdam, New York, Oxford, 1981).

62. Allen, P. B. Theory of thermal relaxation of electrons in metals. *Phys. Rev. Lett.* **59**, 1460–1463 (1987).

63. Chubykalo-Fesenko, O., Nowak, U., Chantrell, R. W. & Garanin, D. Dynamic approach for micromagnetics close to the Curie temperature. *Phys. Rev. B* **74**, 094436 (2006).

64. Kimling, J. *et al.* Ultrafast demagnetization of FePt:Cu thin films and the role of magnetic heat capacity. *Phys. Rev. B* **90**, 224408 (2014).

65. You, W. *et al.* Revealing the Nature of the Ultrafast Magnetic Phase Transition in Ni by Correlating Extreme Ultraviolet Magneto-Optic and Photoemission Spectroscopies. *Phys. Rev. Lett.* **121**, 077204 (2018).

66. Kazantseva, N., Nowak, U., Chantrell, R. W., Hohlfeld, J. & Rebei, A. Slow recovery of the magnetisation after a sub-picosecond heat pulse. *EPL (Europhysics Lett.* **81**, 27004 (2008).

67. Töws, W. & Pastor, G. M. Many-Body Theory of Ultrafast Demagnetization and Angular Momentum Transfer in Ferromagnetic Transition Metals. *Phys. Rev. Lett.* **115**, 217204 (2015).

68. Kirilyuk, A., Kimel, A. V. & Rasing, T. Ultrafast optical manipulation of magnetic order. *Rev. Mod. Phys.* **82**, 2731–2784 (2010).

69. Tengdin, P. *et al.* Critical behavior within 20 fs drives the out-of-equilibrium laser-induced magnetic phase transition in nickel. *Sci. Adv.* **4**, eaap9744 (2018).

70. Mueller, B. Y. & Rethfeld, B. Thermodynamic \textmu T model of ultrafast magnetization dynamics. *Phys. Rev. B* **90**, 144420 (2014).

71. Beens, M., Heremans, J. P., Tserkovnyak, Y. & Duine, R. A. Magnons versus electrons in thermal spin transport through metallic interfaces. *J. Phys. D. Appl. Phys.* **51**, 394002 (2018).

72. Stewart, G. R. Heavy-fermion systems. *Rev. Mod. Phys.* **56**, 755–787 (1984).

73. Hamara, D. *et al.* Ultra-high spin emission from antiferromagnetic FeRh. *Nat. Commun.* **15**, 4958 (2024).

74. Linder, J. & Balatsky, A. V. Odd-frequency superconductivity. *Rev. Mod. Phys.* **91**, 045005 (2019).